%% file: paperFT3.tex
\pdfoutput=1

\documentclass{rspublicp3}
\usepackage{amssymb}

\usepackage[numbers,sort&compress]{natbib}
\usepackage{graphicx}
\usepackage{amsmath}
\usepackage{appendix}
\usepackage{amsthm}

\newtheorem*{thrm}{Competitive H-theorem}
\newtheorem*{thrm2}{Transitive convergence theorem}
\input{tcilatex}

\begin{document}

\title[Complex competitive systems]{Complex competitive systems \\ and \\ competitive thermodynamics. \\
 {\tiny (REVIEW)}  \\ 
{\tiny  Phil. Trans. R. Soc. A 2013, vol.371, No 1982, 20120244} }
\author[A.Y. Klimenko]{A.Y. Klimenko \footnote{email: klimenko@mech.uq.edu.au}}
\affiliation{The University of Queensland, SoMME, QLD 4072, Australia}

\maketitle

\begin{abstract}
{mixing,turbulence, evolving competitive systems, complexity, non-conventional and non-equilibrium thermodynamics}
This publication reviews the framework of abstract competition, which is aimed at studying complex systems with competition in their generic form.  
Although the concept of abstract competition has been derived from a specific field 
--- modelling of mixing in turbulent reacting flows --- this concept is, generally, not attached to a specific phenomenon or application.  
Two classes of competition rules, transitive and intransitive, need to be distinguished. 
Transitive competitions are shown to be consistent (at least qualitatively) with thermodynamic principles, which allows for introduction of special 
competitive thermodynamics.  Competitive systems can thus be characterised by thermodynamic quantities 
(such as competitive entropy and competitive potential), which determine that the predominant direction of evolution of the system 
is directed towards higher competitiveness. There is, however, an important difference: while conventional thermodynamics is constrained by 
its zeroth law and is fundamentally transitive, the transitivity of competitive thermodynamics depends on the transitivity of the competition rules. 
The analogy with conventional thermodynamics weakens as competitive systems become more intransitive, while strongly intransitive competitions can 
display types of behaviour associated with complexity: competitive cooperation and leaping cycles. 
Results of simulations demonstrating complex behaviour in abstract competitions are presented in the electronic supplementary material (ESM).   
\end{abstract}

\section{Introduction}

Thermodynamics occupies a special place among other physical sciences by
postulating irreversibility of the surrounding world as its principal law.
It is not a surprise that thermodynamic principles are often invoked in
relation to various evolutionary processes in which irreversibility plays a
prominent role. While thermodynamics has proven to be successful in
explaining the common trend of moving towards equilibrium states, the
complexity observed in many non-equilibrium phenomena may seem to be
unnecessary if viewed from a thermodynamic perspective. Turbulent fluid
motions, the existence of life forms, the complexities of technological
development and many other processes involving a substantial degree of
coherent behaviour can be mentioned as phenomena that can hardly be
explained by the known trend of entropy to increase in time. On one hand
there are no violations of the first two laws of thermodynamics known to
modern science; on the other hand it is not clear why nature appears to be
more complex than it has to be in order to comply with these laws. Erwin
Schr\"{o}dinger in his famous essay ``What is life?'' \cite{Life1944}
articulated that life forms operate in perfect agreement with the known laws
of physics and must consume exergy (negative entropy) from external sources
to support their existence. Schr\"{o}dinger \cite{Life1944} has also made a
prediction that new laws of nature explaining the complex working of living
organisms will be discovered in the future. This prediction has not come
true yet but, if it ever will, it seems most likely that the unknown laws
will have something to do with studies of complexity.

Although complex systems can be expected to possess some common properties,
complexity represents a notion that is easy to understand intuitively but
difficult to define in rigorous terms \cite{CompEvol1999}. A general
philosophical discussion of complexity might be interesting and informative
but a more quantitative scientific approach to this problem needs a more
specific framework. Very good examples of such frameworks are given by
non-equilibrium thermodynamics \cite{Prig1977}, statistical physics \cite
{ComPT2010}, \textit{algorithmic complexity }(Kolmogorov-Chaitin complexity)
and algorithmic entropy \cite{AlgEntrop2009}, as well as \textit{complex adaptive
systems} \cite{G-M1994,CAS2006}. The present work reviews another framework
-- the framework of \textit{abstract competition} -- which has recently been
suggested to study general principles of complex competitive systems \cite
{K-PS2010,KP2012,K-PS2012}. 

Considering the apparent contradiction between the second law of
thermodynamics and the high degree of organisation present in complex
systems, we are also interested in the possibility (or impossibility) of
effective thermodynamic characterisation of competitive systems. It appears
that competitive systems (at least up to a certain level of complexity) do
allow for a thermodynamic description \cite{K-PS2012}. This conclusion has
profound implications: evolution of competitive systems occurs in a
stochastic manner but in agreement with competitive thermodynamics. It
should be noted, however, that competitive thermodynamics has its
limitations when competition becomes intransitive. As an intransitive system
progresses towards higher complexity associated with competitive cooperation
and cyclic behaviour, the thermodynamic analogy weakens.

The idea of abstract competition was derived from the long-standing
tradition of modelling turbulent combustion \cite
{Pope85,Williams85,Kuznetsov-S-90,Maas1996,KB99,Pope-00,Peters-00,Fox-2003,Heinz-2003,Pitsch2006,Haworth2009,book2011}%
. In these models, it is common to use \textit{Pope particles} (i.e.
notional particles with properties and mixing) \cite{Pope85,KC-PopeFTC}. If
conventional mixing is replaced by \textit{competitive mixing} \cite
{K-PS2010}, these notional particles may bee seen and used as computational
incarnations of generic elements engaged in abstract competition.
Competitive mixing can be deployed to characterise various processes:
turbulent combustion, invasion waves and other related phenomena \cite
{KP2012}. Unlike conventional conservative mixing, competitive mixing can
display complex behaviour with sophisticated interdependencies, which is of
special interest for this review.

The work over this review has stimulated another discovery: abstract
competition seems to be a convergence point for many approaches and ideas
that were developed in very different fields of science and engineering,
sometimes without any apparent relevance to each other. In addition to the
frameworks discussed above, we should mention, \textit{adiabatic
accessibility} \cite{Cara1909,EntOrd2003} and \textit{Gibbs measures} \cite
{GibbsMeasure80}, the \textit{fluctuation theorem} \cite{Ftheor2002} and
variational principles of non-equilibrium thermodynamics \cite
{PrigGlan,MAP1979,Ziegler1983,MEP2006}, economic \textit{utility} \cite
{Mehta1999} and economic cycles \cite{Perez2010}, Eigen's \textit{%
quasispecies model} \cite{Eigen1971} and the \textit{Condorcet paradox} \cite
{Cond1785}, as well as the theory of hydrodynamic turbulence \cite
{Kuznetsov-S-90,Pope-00,MoninY1975}.

\section{Competitive mixing and abstract competition \label{sA4}}

Abstract competition \cite{KlimIMCIC2010} is suggested to study the
principles of competition in their most generic form \cite
{K-PS2010,KP2012,K-PS2012}. Consider a complex competitive system, which has
a large number of autonomous elements engaged in competition with each
other. The evolution of a competitive system involves\ a process of
determining a winner and a loser for competition between any two elements of
the system. The non-conservative properties (i.e. information) of the loser
are lost while the winner duplicates its information into the resource
previously occupied by the loser. The duplication process may involve random
changes, which are customarily called mutations irrespective of the physical
nature of the process. These mutations are predominantly negative or
detrimental but can occasionally deliver a positive outcome. Interaction
between the winner and loser may also involve redistribution of conservative
properties, which is expected to be in favour of the winner (i.e. from the
loser to the winner).

It is easy to see that abstract competition can be represented by a system
of Pope particles, provided conventional conservative mixing is replaced by
competitive mixing. In the present work, the terms ``elements'' and
``particles'' are used synonymously with ``elements'' primarily referring to
competing components of general nature and ``particles'' to their
computational implementations. We mostly focus on the non-conservative
properties, which are most interesting, while limiting our consideration of
conservative properties to the particles themselves (i.e. the number of
particles is preserved by mixing). We also restrict our analysis to
competitive mixing of couples of particles although more complicated schemes
may also be considered if needed.

Let $\mathbf{y}_{p}$ be the set of properties associated with particle $p$.
If particle $p$ appears to be a winner in competition with another particle $%
q,$ we may write $\mathbf{y}_{q}\prec \mathbf{y}_{p}$. On some occasions,
the particles may have the same strength (i.e. $\mathbf{y}_{p}\simeq \mathbf{%
y}_{q}$) and no winner can be determined or the winner has to be selected
randomly. Competitive exchange of information can be illustrated by the
following effective reaction involving the wining particle $p$ and the
losing particle $q$ 
\begin{equation}
\mathbf{y}_{p}+\mathbf{y}_{q}+E\rightarrow \mathbf{y}_{p}+\mathbf{y}%
_{p}^{\prime },\;\;\;\;\mathbf{y}_{p}\succ \mathbf{y}_{q}  \label{reac}
\end{equation}
where $\mathbf{y}_{p}^{\prime }=\mathbf{y}_{p}+\mathbf{m}$ represents a
mutated version of $\mathbf{y}_{p}$, $\mathbf{y}_{p}$ is stronger than $%
\mathbf{y}_{q}$ and $E$ indicates existence of an external source of exergy
that may be needed by these transformations. The mutations $\mathbf{m}$ are
expected to be predominantly negative, which means that $\mathbf{y}_{p}\succ 
\mathbf{y}_{p}^{\prime }$ is much more likely than $\mathbf{y}_{p}\prec 
\mathbf{y}_{p}^{\prime }$.

Abstract competition deals with complex competitive systems (CCS), which
share with complex adaptive systems (CAS, \cite{G-M1994,CAS2006}) their
major premises: 1) working of a complex system is not trivially reducible to
working of its elements and 2) complex systems of different physical origins
should possess some in-depth similarity. CAS and CCS, however, tend to
differ in the other respects.\ The elements of competitive systems compete
rather than adapt and tend to move and mix instead of having fixed
communication links typical for CAS. The long-standing Darwinian tradition
of studying evolutionary systems of high complexity tends to give a higher
priority to adaptation of the system elements to their environments than to
competition between the elements. CAS tend to follow this tradition, while
CCS is focused on competition more than adaptation.

Competitive systems can evolve in a complex manner due to internal
interactions, without any change of external conditions. In these systems,
every competing element is placed in the environment of its competitors. These
interactions may affect every element and at the same time may be affected
by every element participating in competition. Abstract competition focuses
on a joint evolution of a large number of competing elements irrespective of
the physical nature of these elements. If, however, the external conditions
change, a complex competitive system may respond by adapting to new
conditions like any CAS is expected to behave or, if the change is large or
the system is close to its stability limits, by collapsing. \ 

Competitive mixing and abstract competition follow the idea that common
properties of complex systems can be emulated in computational environments.
This revolutionary idea led Eigen to formulate his quasispecies model \cite
{Eigen1971}. Quasispecies evolve in time as determined by their fitness,
i.e. by their specified ability to reproduce themselves. Quasispecies can
compete against each other only indirectly when limitations are imposed on a
common resource.

In abstract competition, the effective fitness of every element is
determined by its competitors and there is no fitness specified as a pre-set
parameter. Performance of an element $\mathbf{y}^{\circ }$ given a
distribution of its competitors $f(\mathbf{y})$ can be quantified by
relative ranking 
\begin{equation}
r(\mathbf{y^{\circ },[}f])=\underset{\infty }{\int }R(\mathbf{y}^{\circ },%
\mathbf{y})f(\mathbf{y})d\mathbf{y},  \label{rank-1}
\end{equation}
where $R(\mathbf{y}^{\circ },\mathbf{y})$ is the antisymmetric index
function taking values $-1$, $1$ or $0$ when the first argument is the
loser, the winner or there is a draw. The relative ranking may range $-1\leq
r\leq 1$ so that $(1+r)/2$ specifies the probability of a win for $\mathbf{y}%
^{\circ }$ while competing with the other elements from the distribution $f(%
\mathbf{y})$. For example, relative ranking becomes $1$ if the element $%
\mathbf{y}^{\circ }$ is stronger than all elements present in the
distribution $f(\mathbf{y})$ or $-1$ if the element $\mathbf{y}^{\circ }$ is
weaker than all elements present in the distribution $f(\mathbf{y})$. As the
competitive system undergoes evolution, the relative ranking of every
element changes since the distribution $f(\mathbf{y})$ changes as well.
Under certain conditions specified by the \textit{Debreu theorem} \cite
{Debreu1954}, which was introduced in economic studies more than half a
century ago, competition can be equivalently characterised by absolute
ranking $r_{\#}(\mathbf{y}^{\circ })=r(\mathbf{y}^{\circ }\mathbf{,}\#)$ so
that 
\begin{equation}
r_{\#}(\mathbf{y}_{p})>r_{\#}(\mathbf{y}_{q})\;\text{ if and only if \ }%
\mathbf{y}_{p}\succ \mathbf{y}_{q}  \label{rabs}
\end{equation}
The absolute ranking determines how a selected element $p$ performs with
respect to another selected element $q$ and it does not depend on $f$ and
does not change when the distribution $f$ evolves. It should be remembered,
however, that the principal conditions of the Debreu theorem ---
transitivity and continuity --- must be satisfied to allow for the
introduction of an absolute ranking. The implications of these conditions
are discussed in the following sections. If fitness is interpreted as a
non-specific indicator of survival and proliferation, it can be reasonably
identified with the absolute ranking.

\section{Competitive thermodynamics}

Competitive mixing can be useful in different applications including
turbulent premixed combustion \cite{KP2012}. The application of competitive
mixing to turbulent premixed combustion follows earlier ideas of Pope and
Anand \cite{Pope-A-85} and results in the equations named after Fisher \cite
{Fisher1937,KPP1937} and Kolmogorov et al \cite{Fisher1937,KPP1937} and
similar to the model suggested by Bray et al. \cite{Bray-L-M-84a}.\ Premixed
combustion is characterised by two major states of the system --- the
reactants and the products. If this process is interpreted as competition,
the products are the winner and the reactants are the loser. The same model
based on competitive mixing can be applied not only to turbulent combustion
but also to a range of other processes such as invasions and simple
epidemics \cite{Mollison1977,KP2012}. It is obvious that transition from
reaction to products in combustion is driven by chemical thermodynamics.
Could there be another kind of thermodynamics that generically favours
winners over losers in the same way as chemical thermodynamics favours
products over reactants? A positive answer to this question leads to the
concept of competitive thermodynamics.

The methodology of thermodynamics requires introduction of entropy. For a
system of Pope particles, the entropy can be introduced in the form \cite
{K-PS2012} 
\begin{equation*}
S=\underset{S_{c}}{\;\underbrace{-n\int_{\infty }f(\mathbf{y})\ln \left( 
\frac{f(\mathbf{y})}{A(\mathbf{y})}\right) d\mathbf{y}}}+\underset{S_{f}}{%
\underbrace{n\int_{\infty }f(\mathbf{y})s(\mathbf{y})d\mathbf{y}}}-\underset{%
S_{e}}{\underbrace{n\ln \left( \frac{n}{e}\right) }}
\end{equation*}
\begin{equation}
\mathbf{=}-n\int_{\infty }f(\mathbf{y})\ln \left( \frac{f(\mathbf{y})}{f_{0}(%
\mathbf{y})}\right) d\mathbf{y+}n\ln \left( \frac{Z}{n}e\right) ,  \label{Sf}
\end{equation}
where 
\begin{equation}
f_{0}(\mathbf{y})=\frac{A(\mathbf{y})}{Z}\exp \left( s(\mathbf{y})\right) 
\label{eq-f0}
\end{equation}
is the equilibrium distribution\footnote{%
Competitive equilibriums do not generally imply achieving the equilibrium
states of conventional thermodynamics. An equilibrated competitive system is
in its stable steady state that is likely consume thermodynamic exergy from
external sources as indicated by $E$ in equation (\ref{reac}).
\par
{}} and 
\begin{equation}
Z=\int_{\infty }A(\mathbf{y})\exp \left( s(\mathbf{y})\right) d\mathbf{y}
\label{eq-Z}
\end{equation}
is the partition function, while $n$ is the number of particles, which is
presumed constant. Equation (\ref{Sf}) is similar to the other definitions
of entropy for particle systems \cite{Dupree1992,Falk2004}. There is a
certain freedom in selecting $A(\mathbf{y}),$ which makes the entropy
definition invariant with respect to replacements of the variables $\mathbf{y%
}$. It might be convenient to simply set\ $A(\mathbf{y})$ to unity, but
linking $A(\mathbf{y})$\ to \textit{a priori probability,} which is related
to steady distributions in absence of the competition and discussed further
in the manuscript, is more justified from the physical perspective. The
entropy $S=S([f])$ is a functional of the distribution $f$ which is presumed
to be normalised. Entropy involves two major components: configuration entropy $%
S_{c},$ which is associated with chaos, and the potential entropy $S_{f}$,
which is associated with the entropy potential $s(\mathbf{y})$ connected to
particle properties $\mathbf{y}$. Since the particles are treated as
indistinguishable (i.e. particle are the same and can be distinguished only
by their properties $\mathbf{y}$), the configurational entropy is reduced by
the particle exchange term $S_{e}=\ln (n!)\approx n\ln (n/e)$, where $e=\exp
(1)$. The introduced entropy can be interpreted as being similar to the free
entropy of conventional thermodynamics ($S_{G}=-G/T$ where $G$ is the free
energy, due to Gibbs or Helmholtz).

We may or may not know the exact mechanism behind higher competitiveness of
some states as compared to the other states but, in any case, the inequality 
$r_{\#}(\mathbf{y}_{1})>r_{\#}(\mathbf{y}_{2})$ indicates that nature
prefers state $\mathbf{y}_{1}$ to state $\mathbf{y}_{2}$. A thermodynamic
expression for the same property is $s(\mathbf{y}_{1})>s(\mathbf{y}_{2})$.
Hence, entropy potential $s$ and absolute ranking $r_{\#}$ are linked to
each other $s=s(r_{\#})$. Higher ranking $r_{\#}$ and higher entropy
potential $s$ recognise a greater affinity of nature towards these states,
while the physical reasons responsible for this affinity may differ.

The connection between transitive continuous ordering of states by a ranking
function and thermodynamic entropy is known in conventional thermodynamics
under the name of adiabatic accessibility. This principle was introduced by
Caratheodory \cite{Cara1909} and recently used by Lieb and Yngvason \cite
{EntOrd2003} to successfully deduce the whole structure of conventional
thermodynamics from this principle\footnote{ A similar approach, 
albeit based on a different quantity  -- 
{\it adiabatic availability}, has been previously developed by Gyftopoulos and Beretta
\cite{Beretta1991}}. It is useful to note that similar
methods have been under development in theoretical physics and mathematical
economics for more than half a century without any knowledge or interaction
between these fields. The similarity between introducing economic utility
and defining entropy on the basis of adiabatic accessibility was noticed
first by Candeal et. al. \cite{Mehta2001}, who called it ``astonishing''.
This similarity seems to indicate that the concept of entropy should be also
relevant to economic studies \cite{K-PS2012}.

While the physical arguments in favour of connecting the entropy potential
to the ranking seem convincing, constructing competitive thermodynamics
requires the competitive analogue of the second law to be established.
Indeed, if a competitive system evolves in isolation, its entropy $S$ should
be defined so that it always remains a non-decreasing function of time. This
appears to be correct for simpler cases but as simulations progress towards
cases with greater complexity the thermodynamic analogy weakens.

\section{Gibbs mutations and competitive H-theorem}

In this section we consider a special type of mutations --- the \textit{%
Gibbs mutations }--- which ensure maximal consistency with thermodynamic
principles. The results are applicable to both transitive and intransitive
competitions. A general definition of Gibbs mutations is given Ref. \cite
{K-PS2012}. Here we outline major properties of Gibbs mutations and discuss
their relevance to Markov properties and Gibbs measures \cite{GibbsMeasure80}%
.

The Gibbs mutations are defined as strictly non-positive 
\begin{equation}
\mathbf{y}_{p}^{\prime }=\mathbf{y}_{p}+\mathbf{m\preccurlyeq y}_{p}
\end{equation}
We can consider Gibbs mutations as being represented by a sequence of a
large number of small steps directed towards states with reduced
competitiveness. The probability of each subsequent step is independent of
the proceeding steps, which essentially represents a Markov property. If a
step does not occur, the whole process is terminated and the final point of
the mutation is reached. The probability distribution of these mutations can
be interpreted as a Gibbs measure \cite{K-PS2012}.

Equation (\ref{Sf}) indicates that entropy achieves it maximum $S\rightarrow
n\ln (eZ/n)$ when the distribution $f$ approaches its equilibrium $f_{0}$.
Evolution of competitive systems appears to be consistent with the following
theorem:

\begin{thrm}
An isolated competitive system with Gibbs mutations evolves in a way that
its entropy $S$ increases in time until $S$ reaches its maximal value at the
equilibrium.
\end{thrm}

The equations governing the evolution of competitive systems and proof of
the competitive H-theorem can be found in Ref. \cite{K-PS2012}. This theorem
indicates consistency of the introduced entropy with the principles of
thermodynamics and is valid for both transitive and intransitive competitions.

The usefulness of the thermodynamic description can be illustrated by the
following consideration: assume that two competitive systems with a fixed
total number of particles $n=n_{1}+n_{2}=\func{const}$ are brought in
contact with each other; then the total entropy $S=S_{1}+S_{2}$ must reach
its maximum when the equilibrium between the systems is established after
particle exchange. Hence, the values $\chi _{1}=\partial S_{1}/\partial
n_{1}=\ln (Z_{1}/n_{1})$ and $\chi _{2}=\partial S_{2}/\partial n_{2}=\ln
(Z_{2}/n_{2})$ should be the same in the equilibrium. These values $\chi
_{I},$ $I=1,2$ are called \textit{competitive potentials }and represent
quantities analogous to the chemical potential of conventional
thermodynamics taken with negative sign. A non-zero difference between
competitive potentials $\chi _{1}-\chi _{2}$ points to the direction of flow
of the resources when two competitive systems are brought into contact.

The Great American Exchange, which took place around 3 million years ago
when the Isthmus of Panama connecting two Americas was formed, may serve as
an example \cite{GAI1982}. Before the Isthmus was formed the biosystem of
each continent wwas, presumably, close to quasi-equilibrium (see the next
section). The exchange induced significant and complex changes in the
biosystems. While each of these changes may have its specific cause or
explanation, the methodology of thermodynamics neglects the details and
recognises the overall trend behind a large number of seemingly unrelated
events. On average, the North American animals performed better than their
South American counterparts consistently in both native and immigrant
conditions. We may infer that, at the time of the connection, the fauna of
North America had somewhat higher competitive potential than the fauna of
South America (which is difficult to explain by adaptation).

\section{Transitive competition}

When mutations deviate from Gibbs mutations, the behavior of the competitive
system becomes quite different for transitive and intransitive competitions.
Competition is deemed transitive when the following statement 
\begin{equation}
\text{ if \ }\mathbf{y}_{1}\preccurlyeq \mathbf{y}_{2}\text{ and }\mathbf{y}%
_{2}\preccurlyeq \mathbf{y}_{3}\;\text{\textbf{\ }then \ }\mathbf{y}%
_{1}\preccurlyeq \mathbf{y}_{3}
\end{equation}
is valid for any three states indexed here by 1, 2 and 3. Presuming
continuity of the competition rules, transitive competition can be
characterised by absolute ranking (\ref{rabs}). The particle that has
the highest absolute rank in the distribution $r_{\ast }=r_{\#}(\mathbf{y}%
_{\ast })$ is called the leading particle and its location is denoted by $%
\mathbf{y}_{\ast }$.

In transitive competition with negative mutations the leading particle can
not lose to any other particle and can not be overtaken by any other
particle -- hence $\mathbf{y}_{\ast }$ remains constant and, according to
the H-theorem, the distribution $f$ converges to its equilibrium state $%
f_{0}(\mathbf{y},\mathbf{y}_{\ast })$. However, existence of positive
mutations results in the occasional particle jumping in front of the leader. The
overtaking particle then becomes a new leader and $\mathbf{y}_{\ast }(t)$ is
an increasing function of time. If positive mutations are small and
infrequent, the distribution $f$ remains close to quasi-equilibrium $f_{0}(%
\mathbf{y},\mathbf{y}_{\ast }(t))$ but progresses towards higher ranks as
illustrated in Figure 1a. This process, which is named \textit{competitive
escalation}, results in increase of $S$ in time and is perfectly consistent
with competitive thermodynamics.

The behavior of competitive systems is similar to that of conventional
thermodynamic systems and generally consistent with Prigogine's minimal
entropy production principle \cite{PrigGlan}. The\ generation of entropy $S$
decreases as the distribution $f(\mathbf{y},\mathbf{y}_{\ast })$ rapidly
approaches its quasi-equilibrium state $f_{0}(\mathbf{y},\mathbf{y}_{\ast })$
and then takes a relatively small but positive value as $r_{\mathbf{\#}}(%
\mathbf{y}_{\ast })$ increases due to competitive escalation. The rate of
competitive escalation is evaluated in and is determined by the average
magnitude of positive mutations, which is linked by the \textit{fluctuation
theorem} \cite{Ftheor2002} to the gradients of \textit{a priori entropy} $%
\hat{s}(\mathbf{y})$ \cite{K-PS2012}. The a priori entropy $\hat{s}$ is
related to the steady distribution of particles in absence of competition $%
\hat{f}_{0}\sim \exp (\hat{s})$, which is named \textit{a priori probability}%
, and should be distinguished from the entropy potential of the competition $%
s(\mathbf{y}).$ The entropies $\hat{s}(\mathbf{y})$ and $s(\mathbf{y})$
increase in opposite directions since more competitive states with higher $s(%
\mathbf{y})$ are relatively rare and have smaller a priori probability $\hat{%
f}_{0}$. The applicability of the maximal entropy production principle (MEP) 
\cite{MAP2005,MEP2006} (representing a generalisation of the principle
introduced by Ziegler \cite{Ziegler1983}) to competitive systems remains an
open question. Note that there is no contradiction between the principles of
Prigogine and Ziegler as they are formulated for different constraints \cite
{MEP2006}. 

If mutations significantly deviate from Gibbs mutations, the thermodynamic
analogy weakens but, as long as competition remains transitive, the
behaviour of competitive systems is qualitatively similar to that with Gibbs
mutations:

\begin{thrm2}
An isolated system with transitive competition and non-positive mutations
converges to a unique equilibrium state in which the entropy $S$ reaches its
maximal value (for a fixed location $\mathbf{y}_{\ast }$ of the leading
element).
\end{thrm2}

This theorem, which is proven in Ref. \cite{K-PS2012} under certain
mathematical constraints, does not guarantee monotonic increase of entropy
and existence of the detailed balance at equilibrium (these are guaranteed
by the competitive H-theorem). Deviations from Gibbs mutations, however, do
not affect the qualitative behaviour of a system with transitive
competition: the system first rapidly approaches its quasi-equilibrium state
and, provided infrequent positive mutations are present, then escalates
slowly towards more competitive states and larger $S$. Competitive
escalation, which involves gradual increase of competitiveness in
competitive systems, is generally quite realistic but can not explain more
sophisticated patterns of behaviour observed in the real world. An evolving
transitive system may diverge as shown in Figure 1b but divergence by itself
does not remove the restrictions of transitivity imposed on each of the
separated subsystems. Competitive escalation cannot continue indefinitely:
sooner or later the competitive system should encounter restrictions
associated with resource limitations, laws of physics, etc. and at this
point any further development would be terminated. Evolutionary processes
may still act to reduce the magnitude of mutations resulting in a slight
additional increase in competitiveness (assuming that the magnitude of
mutations is allowed to mutate --- note that reduction of the magnitude of
mutations is the only possible way for further increase in particle
competitiveness in conditions when the leading particle has the maximal
ranking permitted in the system). At the end, the system enters into the
state of global equilibrium and cannot change, except if the external
conditions are altered. Unless these external conditions become increasingly
complex, the transitive system is likely to respond to this alteration by a
relatively small adjustment and any significant increase in complexity
remains unlikely in these conditions. A very large alteration of the
environment may, of course, destroy the system. Consequently, complexity
emerging within competitive systems should be associated with overcoming the
major constraint imposed by transitive competition rules, i.e. with
intransitivity.

\section{Intransitive competition}

Competition is deemed intransitive when there exist at least three states $%
\mathbf{y}_{1}$, $\mathbf{y}_{2}$ and $\mathbf{y}_{3}$ such that 
\begin{equation}
\mathbf{y}_{1}\preccurlyeq \mathbf{y}_{2}\preccurlyeq \mathbf{y}_{3}\mathbf{%
\prec y}_{1}  \label{int3}
\end{equation}
For a long time, since the days of the French revolution when intransitivity
was first studied by outstanding mathematician, philosopher and humanist
marquis de Condorcet \cite{Cond1785}, intransitivity was mostly viewed as
something unwanted or illogical \cite{Intrans1964}. Arrow's theorem \cite
{Arrow}, which is well-known is social sciences, indicates that
intransitivity may even pose a problem to choice in democratic elections.
Intransitivities have nevertheless become more philosophically accepted in
recent times \cite{Intrans1993} and are now commonly used in physics \cite
{Intrans2011}, biology \cite{BioInt2000} as well as in social studies \cite
{Rank2004}. Recently published works \cite{K-PS2010,K-PS2012} indicate the
importance of intransitivity for evolution of complex systems, although
intransitivity may indeed have some ``unwanted side effects''.

Although in the case of Gibbs mutations intransitivity does not violate the
thermodynamic analogy, the effects of any deviation from Gibbs mutations
combined with intransitivity can produce, depending on the conditions, very
diverse patterns of behaviour. Intransitivity conventionally refers to any
violation of transitivity, irrespective of intensity of these violations.
Intransitive competition rules may effectively combine some transitive and
intransitive properties. An example of a globally intransitive system that
remains locally transitive is shown in Figure 1c. The arrow indicates the
direction of increasing strength; this direction forms a circle so that $k$
locations forming an intransitive loop $\mathbf{y}_{1}\mathbf{\prec y}_{2}%
\mathbf{\prec }...\mathbf{\prec \mathbf{y}_{k}\prec y}_{1}$ can be easily
nominated (the strengths of any two particles are compared in the direction
of the shortest arc connecting these particles). We note that the absolute
ranking $r_{\mathbf{\#}}\mathbf{(y})$ does not exist as a conventional
function and $r_{\#}\mathbf{(y})$ becomes multivalued $r_{\mathbf{\#}}%
\mathbf{(y}_{1})<$ $...$ $<r_{\mathbf{\#}}\mathbf{(y}_{k})<r_{\mathbf{\#}}%
\mathbf{(y}_{1}).$ Clearly, evolution of this system would be very similar
to transitive competitive escalation although the whole process becomes
cyclic and the same distribution of particles (such as shown in Figure 1c)
would be repeated periodically. Competition entropy $s(\mathbf{y})$ can be
introduced locally in the same way as $s(\mathbf{y})$ is introduced for
transitive competition but, on a large scale, the thermodynamic quantities
become multivalued $s=s(r_{\mathbf{\#}}\mathbf{(y}))$. We may assume for
simplicity that mutations are close to Gibbs mutations at every location and
nevertheless find three different systems $I=1,2,3$ with competition
potentials $\chi _{1},$ $\chi _{2}$ and $\chi _{3}$ that are not transitive
and symbolically satisfy the inequalities $\chi _{1}<\chi _{2}<\chi
_{3}<\chi _{1}$. This, indeed, must be a very unusual thermodynamics. In
conventional thermodynamics, temperatures and chemical potentials are
restricted by the zeroth law of thermodynamics and are fundamentally
transitive.

Another example, presented in Figure 1d, shows a competition that has a
transitive direction of increasing strength along the $y_{1}$ axis combined
with cyclically intransitive rules (the same as in the previous example) on
every plane $y_{1}=\func{const}.$ This system would undergo a spiral
evolution including translational competitive escalation in the direction of
transitive increase of the competitive strength and cyclic motions along the
plane of constant $y_{1}$. The system escalates in the transitive direction
until the maximal possible value of $y_{1}$ is reached. At this moment
transitive evolution in the direction of $y_{1}$ is terminated while
intransitive evolution may continue its cycles indefinitely.

The main feature of intransitive competitions is the absence of clear
signposts for being stronger and being weaker. The weakest particle, if left
in isolation behind the distribution circulating in Figure 1c, would
eventually become the strongest particle in the set. An improvement in a
transitive system is always an improvement while a direction that seems to
improve competitiveness of an intransitive system in the present conditions
may in fact reduce its competitiveness when the overall distribution changes
over time. In intransitive conditions, decisions that seem very
reasonable in the current situation may\ appear to be detrimental in the
long run (and vice versa). For example, investments in ``dot-com'' companies
were seen as quite profitable in the 1990s but the same investments would be
viewed very differently in the 2000s. A strategy in an intransitive system
needs to not only be evaluated with respect to the current criteria of
competitiveness, but also examined against the changes that are likely to be
introduced as the system evolves. Intransitivity brings more diverse cyclic
patterns of behaviour into competitive systems, while real-world
competitions can be controlled by very complicated combinations of
transitive and intransitive rules. The overall trend in evolution of
competitive systems is the elimination of transitively weak elements,
resulting in predominance of the intransitive rules in the competition (since
intransitive weaknesses give better chances of survival as compared to
transitive weaknesses).

\section{Intransitivity and complexity}

The systems considered in the previous section are intransitive but display
many features of transitive competitions. This allows for application of
competitive thermodynamics, which, at this point, diverges from conventional
thermodynamics and involves intransitivities. We now move to consider
competitions with strongly intransitive rules, where intransitive triplets (%
\ref{int3}) can be found in every open neighbourhood of any location $%
\mathbf{y}$. In these competitions, the absolute ranking cannot exist even
locally. Particle strength may still be evaluated in terms of the relative
ranking $r(\mathbf{y,}[f_{r}])$ but this ranking is highly dependent on
selection of a suitable reference distribution $f_{r}$: improvement of
ranking of a particle with respect to one reference distribution may reduce
the ranking of the same particle with respect to another reference
distribution. In intransitive systems, assessment of progress and regress
may strongly depend on the observer's perspective.

The most interesting point in examining evolution of competitive systems is
the possibility or impossibility of a complex behaviour, which implies a
sophisticated coordination of properties of many elements in a way that
apparently involves or resembles formation of structures and a joint action
of elements within each structure. The usefulness of applying the thermodynamic
analogy to strongly intransitive competitions seems questionable \cite
{K-PS2012}. The diversity of particle distributions associated with complexity
implies localisation of these distributions and, consequently, localisation
of competitive interactions in physical space. Abstract competition views
intransitivity and localisation as necessarily conditions for complex
behaviour, although these conditions may be insufficient to guarantee
complexity (a system combining intransitivity with Gibbs mutations, which
enforce relatively simple relaxation to equilibrium, may serve as a
counterexample). From practical perspective, however, a sufficiently large
system with strong intransitivity and localisation of competitive
interactions in physical space seems to have a non-vanishing probability of
developing complex behaviour. Indeed, even the most simple competitive
systems that still possess the properties of strong intransitivity and
localisation display clear signs of developing complexity under appropriate
conditions \cite{K-PS2010}.

Complex behaviour, which is associated with strong intransitivity and
localisation, is accompanied by a number of indicators that have been
observed in computer simulations of strongly intransitive abstract
competition \cite{K-PS2010}. It should be stressed that mutations used in
these simulations are purely random and do not have any purpose or
coordination between particles. The results of simulations presented in the
electronic supplementary material (ESM) indicate diverse patterns of
behaviour, which are associated with complexity and discussed in the
following subsections.

\subsection{Competitive cooperation}

\textit{Competitive cooperation }is the formation of structures accompanied
by coordination of properties and reduction of competition intensity within
the structures (as compared to intensity of competition between the
structures). Survival of element properties within the structure is strongly
correlated with survival of the whole structure. Complex cooperative
behaviour occurs when competitive mixing is strongly intransitive and
localised in physical space \cite{K-PS2010,K-PS2012}. The computer
simulations of abstract competition presented in ESM involve one of the
simplest possible systems that possess the properties of strong
intransitivity and localisation of mixing and, indeed, these simulations
display clear cooperative behaviour. The simulations indicate the following
mechanisms of competitive cooperation: 1) particles have more similar
properties within each structure as compared to larger differences of
particle properties between the structures and 2) each structure resembles a
pyramid where particles tend to compete against particles of similar
relative ranks within the structure (i.e. leaders against leaders and
low-rankers against low-rankers).

Coordination of properties between elements and formation of structures is
the key indicator that distinguishes partially ordered from completely
chaotic types of behaviour. Cooperation thus increases complexity and, if
the size of the system allows, inevitably leads to even more complex
behaviours. Indeed, the formed structures may be considered in the framework
of abstract competition as new elements (i.e. superelements) competing
according to the integrated competition rules. These integrated competition
rules are inherently intransitive and interactions between structures are
localised in physical space (the \textit{competitive Condorcet paradox} \cite
{K-PS2012} indicates that the integrated competition rules are likely to be
intransitive even if the underlying competition rules between primary
elements are transitive). Hence the structures are likely to form groups of
structures coordinating the properties of structures and the overall picture
presented by abstract competition becomes more and more complicated. If the
system is sufficiently large, the hierarchy of structures can grow to become
very complicated. Competitive cooperation, which in most cases is seen as a
very positive factor, may nevertheless have unwanted side effects: the
reduction of competition associated with competitive cooperation can be
accompanied by degradations.

\subsection{Competitive degradation}

\textit{Competitive degradation} is reduction of competitiveness that may
appear in intransitive competitions under certain conditions. Competitive
degradation is the process opposite to competitive escalation and,
generally, is not consistent with competitive thermodynamics (see ref. \cite
{K-PS2012} for further discussion).

Owing to the relativistic nature of competitiveness quantified in intransitive
conditions by relative ranking $r(\mathbf{y,}[f_{r}])$, the presence of
competitive degradations may or may not be obvious. While degradations are
unambiguously present in simulations shown in the ESM, degradations can also
be latent and might be superficially seen as escalations, especially
when considered from a certain fixed perspective (i.e. using specific poorly
selected $f_{r}$ in $r(\mathbf{y,}[f_{r}])$). \ A competition process may
result in increase of competitiveness with respect to one parameter, say $%
y^{(1)},$ and decrease of competitiveness with respect to another parameter,
say $y^{(2)}.$ In transitive competitions, the former always outweighs the
latter and the overall absolute ranking must increase. Intransitivity makes
the outcomes much less certain. Increase in $y^{(1)}$ may seem to be\ a
competitive escalation while a minor decrease in $y^{(2)}$ may not even be
noticed. However, if and when $y^{(2)}$ falls below a certain critical
level, the dominant structure may become unstable with respect to large disturbances, 
which can be created by rare mutations, and ultimately collapse.

Competitive degradations should be clearly distinguished from \textit{%
erosive degradations:} the former are the results of strict compliance with
the competition rules while the latter appear due to various imperfections.
Erosive degradations may exist in any competition while competitive
degradations require intransitivity. In our analysis of competitions, we
have assumed that information has eternal properties: that is, once obtained
and not overridden in competition, information can exist forever. In the
real world, however, information may be eroded and lost. Another cause of
erosions may be imperfections in the competition process inducing random
outcomes so that the distributions become more and more influenced by the a
priori probability. Competitive degradation is different: it is a result of
``perfect'' competition when at every step the strongest always wins but,
nevertheless, the overall competitiveness declines due to interactions of
intransitivity and mutations. This process may seem abnormal but computer
simulations \cite{K-PS2010} show that it can exist under conditions of
strong intransitivity and localisation.

Let us consider two examples. 1) Company managers become bureaucratic,
following all formal procedures but stalling initiative and, as a result,
the quality of the company's products declines. This is competitive
degradation. 2) Old technical drawings lose information due to paper erosion
and, because of this, the company can not sustain the quality of its
products. This is erosive degradation.

It is quite obvious that erosive degradations are much easier to detect and
prevent (at least theoretically) as compared to competitive degradations,
whose causes are more sophisticated. It seems that competitive degradations
represent a possible side effect of competitive cooperation (see ESM).
Cooperations are accompanied by a reduction of competition that, in turn,
may be followed by degradations.

\subsection{The leaping cycle}

The \textit{leaping cycle} is a special type of cyclic behaviour that is
characterised by the emergence of a new structure, followed by its rapid
growth and leapfrogging into a strong or even dominant position, stable
strength or domination, decline, and a complete collapse or restricted
(niche) existence. The leaping cycle differs from ordinary cycles (such as
depicted in Figure\ 1c): leaping cycles are associated with competitive
degradation and result in replacement of the old structures by new ones,
while ordinary cycles are caused by local competitive escalation in
conjunction with cyclic intransitivity and result in periodic repetitions of
the same states.

Practically, a complicated cycle can possess features of both ordinary and
leaping cycles while degradations and escalations may not be easily
distinguishable. The term leaping cycle refers to a generic form of the
cycle, such as observed in simulations presented in ESM. In different
disciplines, cyclic behaviours similar to the leaping cycle are known under
different names \cite{KlimIMCIC2010}. Technological waves with durations of
around 50 years may serve as an example of cyclic behaviour. These cycles
were discovered by Kondratiev \cite{Kond1925} nearly 100 years ago as
long-term oscillations in economic growth --- according to this
understanding the Kondratiev technological waves represent ordinary cycles.
Another interpretation of the technological cycles involving irruption,
rapid growth, maturity and decline is recently given by Perez \cite
{Perez2010}. This interpretation resembles the pattern of a leaping cycle
much more than that of an ordinary cycle.

\section{Intransitivity in turbulent flows}

The presented treatment of competitive systems is generic and can be applied
to systems of different physical nature. In this section, as at the
beginning of this review, we consider the application of competition
principles to turbulent flows. The primary goal of this section is not in
suggesting a new approach for treatment of turbulent flows but in looking at
existing approaches from a different perspective based on principles of
abstract competition.

Eddies of different sizes form the turbulent energy cascade \cite
{Batchelor-53,MoninY1975,Kuznetsov-S-90,Pope-00,Tsinober2009}. The
development and breakdown of turbulent eddies can, in principle, be
interpreted as competition between the eddies This interpretation, however,
must remain qualitative as eddies in turbulent flow tend to erode quickly
and cannot be defined as fully autonomous objects. A degree of complexity
may be present in turbulent flows but it must be restricted by the high rate
of eddy erosion. ``Eddy competition'' should be clearly distinguished from a
more rigorous application of competitive mixing to turbulent reacting flows.
Since the states of chemical composition are well-defined and evolve due to
the well-defined combination of mixing, transport and reactions, reacting
flows are much more suitable for quantitative application of competitive
mixing \cite{KP2012}.

Turbulence is a complex phenomenon which can provide examples of
intransitive behaviour. Here we refer to three-dimensional turbulence since
the properties of two-dimensional turbulence are quite different.
Specifically, the inverse cascade of two-dimensional turbulence preserves
energy and eddy vorticity, which makes this phenomenon amendable to
thermodynamic description \cite{Som2005}. The eddies interact in
three-dimensional turbulence, which is interpreted here as ``competition''
of eddies. It is interesting that winners and losers can be defined for
interactions of eddies in turbulent flows in different ways. Energy is
transferred from larger to smaller eddies, hence from the perspective of
conservative properties (i.e. energy) smaller eddies are the winners.
Considering non-conservative properties, it is larger eddies that
statistically control the structure and intensity of the smaller eddies and
the larger eddies are thus the winners. In the present consideration, we
focus on information flows and treat larger eddies as winners and smaller
eddies as losers.

Figure 2 shows the directions of control from the mean flow through
instabilities first to large and then to smaller and smaller eddies. This
figure follows Kuznetsov's theory of energy transport from the mean shear
flow into large-scale oscillations of fully developed turbulence through
growth of instabilities (this theory is developed for turbulent shear flows
with a mean velocity profile having at least one inflection point \cite
{Kuz1989}). The scheme, which is shown in Figure 2 by solid arrows, is
transitive. In transitive competitions, particles with higher absolute ranks
are not directly affected by particles with lower absolute ranks, since the
former are always the winners and the latter are always the losers.
According to the transitive scheme, turbulence would behave in a predictable
manner, because the problem can be solved at larger scales without emulating
the turbulent motions at smaller scales. Many of the empirical models of
turbulence are based on this assumption, although some methods (i.g. the
probability density function methods) are aimed at simulating stochastic
behaviour at smaller scales \cite{Pope-00}. A number of theories dealing
with the energy dissipation cascade implicitly incorporate transitivity by
assuming stochastic control of turbulent motions at larger scales over
turbulent motions at smaller scales \cite{MoninY1975}. While these models
and theories often achieve a reasonable degree of accuracy, turbulence has
proven to be a more complex phenomenon than is predicted by any existing
theory or model. Turbulence is commonly believed to be the last great
unsolved problem of classical physics (see the excellent collection of
quotations about turbulence in Ref.\cite{Tsinober2009}).

The arguments presented in this review link complexity to intransitivity. Is
this link relevant to turbulence? Intransitive exchange between the Reynolds
stress components in shear flows \cite{K-PS2012} presents an example of
intransitivity in turbulence, but here we are interested in the turbulent
eddy cascade. If the eddies are deemed to ``compete'' against each other for
control within the cascade, intransitivity should be understood as some
degree of control of smaller eddies over motions at larger scales. This
reverse control is perhaps not as strong as the direct control of larger
eddies over smaller eddies but its existence would promote the smaller
eddies from the status of mere followers to the status of (partially)
independent players. This makes evaluation of motion at larger scales
dependent on details of the stochastic events at smaller scales, which poses
a problem for any models that are restricted to consideration of large
scales. Figure 2 shows the direction of reverse control from smaller to
larger scales by the dashed arrow. The reverse control, which may involve
the influence of turbulent viscosity on the rate of growth of the
instabilities in the mean flow and other mechanisms, introduces
intransitivity into the turbulent cascade.

\section{Conclusions}

Although abstract competition was derived from ideas used in modelling of
mixing in turbulent reacting flows, it pertains to study of competition
principles in their most generic form. Competitive systems can be complex
(CCS) and have a number of features common with complex adaptive systems
(CAS). Abstract competition involves exchanging properties between competing
elements and, hence, can be seen as a form of mixing (i.e. competitive
mixing). This exchange distinguishes winners and losers and discriminates
against the losers. The winners and losers are selected on the basis of
element properties as interpreted by a set of competition rules. The degree
of transitivity (or intransitivity) of the competition rules is the one of
the main factors affecting the behaviours of competitive systems.

Transitive competitions result in competitive escalations, which are
accompanied by gradual increase in competitiveness of the system, or in
achieving a global equilibrium state that represents the final state of
evolution of the system. The analogy with conventional thermodynamics is
especially strong in the case of Gibbs mutations. Transitive evolutions are
consistent with a thermodynamic description, at least qualitatively: the
competitive entropy tends to stay the same or increase. From the perspective
of competitive thermodynamics, the phenomenon of achieving higher
competitiveness in transitive competitive systems is natural, in the same
way that a formation of crystalline structures under appropriate conditions
is natural in conventional thermodynamics. The complexity of transitive
competitions, however, remains restricted and falls significantly short of
explaining the complexity observed in nature.

As competition becomes more and more intransitive the analogy with
conventional thermodynamics, which is fundamentally transitive, weakens.
Intransitivity leads to a situation where thermodynamic functions become
multivalued functions of the state of the system; for example, competitive
potentials may satisfy the symbolic inequality $\chi _{1}<\chi _{2}<\chi
_{3}<\chi _{1}$. Needless to say, similar inequalities are impossible for
temperatures or chemical potentials in conventional thermodynamics. This
type of intransitivity allows for endless cyclic oscillations in competitive
systems.

Complex types of behaviour are associated with stronger intransitivity, when
intransitive triples can be found in the vicinity of any point, and with
localisation of competitions in physical space. Although strong
intransitivity and locasation may not necessarily be sufficient for
development of complexity, the emergence of complexity becomes rather likely
under these conditions since even the relatively simple system presented in
ESM (which is essentially the simplest possible system with strong
intransitivity and localisation) displays complex behaviour. The
thermodynamic analogy tends to become inapplicable as the system begins to
display more complex patterns of behaviour, among which we outline
competitive cooperation, competitive degradation and the leaping cycle.
Competitive cooperation is formation of structures with a reduced level of
competition within each structure. Competing particles or elements survive
competition through a coordinated effort rather than individually.
Competitive degradation, which seems to be a side effect of competitive
cooperation, is an ``abnormal'' outcome of competition resulting in reduced
competitiveness. Evolution in intransitive systems can be highly non-linear
involving emergence, a leap forward to a dominant position, stagnation,
degradation and, possibly, collapse. In abstract competition this type of
behaviour is called a leaping cycle, although it may be related to cycles
known under different names in different disciplines.

We conclude by noting the natural trend of progressing towards complexity in
competitive systems. Realistic competitions are likely to include a mixture
of transitive and intransitive rules. While transitive features tend to
dominate at the initial stages, elimination of transitively weak elements
results in competitions being restricted to regions where intransitivity
prevails. Since localisation in physical space is dictated by the laws of
physics, the combination of intransitivity and localisation creates the
conditions needed for competitive cooperation and formation of structures.
Assuming that the overall size of the system is extremely large, the
principles of abstract competition can now be applied to these structures
rather than to the elements. The rules for competition between the
structures should also be intransitive, while their interactions remain
localised in physical space. This leads to cooperative behaviour between the
structures and formation of superstructures. Recursive application of the
principles of abstract competition to superstructures at different levels
opens the possibility of building competitive hierarchies of unlimited
complexity.

\begin{acknowledgements}
The author thanks D.N.P. Murthy for numerous discussions and S.B.Pope for
insightful comments and suggestions. Suggestions made by the lead Editor of the
issue and by the anonymous reviewers are also appreciated by the author. The
author thanks D.N. Saulov and D.A. Klimenko for useful comments and
assistance in preparation of the manuscript. The part of this work related
to development of particle methods in application to reacting flows is
supported by the Australian Research Council.
\end{acknowledgements}

\newpage

\begin{figure}[ht] 
\caption{Evolutions in competitive systems: a) competitive escalation in a non-preferential transitive competition; 
b) competitive escalation in transitive competition with particle isolations causing divergence of distributions; 
c) evolution in a system with cyclic intransitivity;
 d) spiral evolution in a more complex system combining transitive and intransitive properties. 
 The arrows with the oval-shaped starting points show the direction of competitive escalation and increasing ranking. } 
\begin{center} 
\includegraphics[width=8cm]{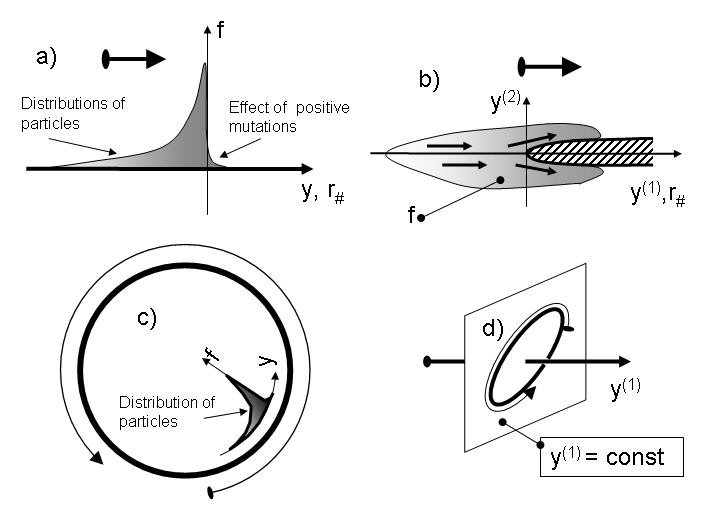}  
\label{fig_1}  
\end{center}
\end{figure}

\begin{figure}[hb] 
\caption{Presence of intransitivity in the turbulent cascade.} 
\begin{center} 
\includegraphics[width=8cm]{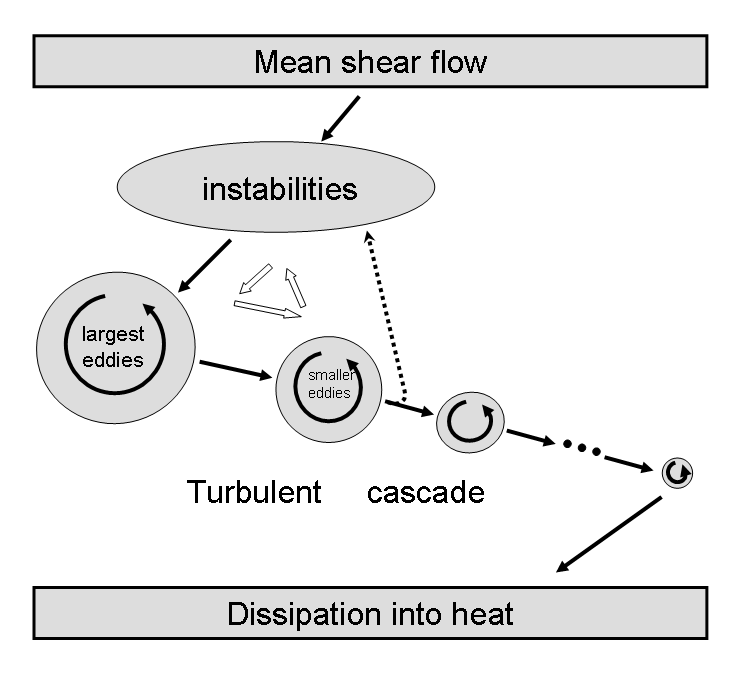}  
\label{fig_2}  
\end{center}
\end{figure}

\include{supFT}

\end{document}

%% file: supFT.tex




\input{ins1sup}

Collection of snapshots of the physical domain obtained from the ``Long
Run'' simulation, which involves 64 000 Pope particles and lasts over 200
000 time steps. The numbers above the snapshots indicate the time step.

\pagebreak

\setcounter{section}{0}
\renewcommand{\thesection}{\Alph{section}} 

\section{Simulations of abstract competition}

The figures and videos presented in this supplement have been
obtained from computer simulations of a complex competitive system within
the framework of abstract competition. These simulations produce the
impression of coordinated efforts of different groups of particles that seem
to be directed at establishing domination of these groups over the whole
domain. Although we may use natural terminology derived from this intuitive
interpretation, it is important to remember that this interpretation is only
an illusion. The particles are fully controlled by relatively simple
competition rules followed by mutations, which are kept completely random
and uncorrelated for different particles. Other than that, there is no
programming coordination of particle behaviour of any kind. Although there
is no purpose in particle interactions, the system displays a number of
features associated with complex behaviour, which we intuitively identify
with other complex systems that do possess a purpose or a plan. As explained
in the printed part of the manuscript, these features are associated with
intransitive localalised competitions and represent generic patterns of
behaviour of competing systems. These patterns involve:

\begin{itemize}
\item  \textit{Competitive cooperation }

\item  \textit{Competitive degradation }

\item  \textit{Leaping cycles }
\end{itemize}

The simulations presented here are performed using 64000 Pope particles
during a ``Long Run'' involving $2\times 10^{5}$ time steps. These particles
move in 2-dimensional physical space according to an Ornstein-Uhlenbeck
vector process (each particle moves independently) and exchange their
properties through competitive mixing between particles. Competitive mixing
is localised in the physical space, i.e. only particles close in the
physical space are allowed to compete directly. The particle properties are
represented by the vector $\mathbf{y}=(y^{(1)},$ $y^{(2)},$ $y^{(3)})$,
which satisfies the conservation constraint 
\begin{equation}
y^{(1)}+y^{(2)}+y^{(3)}=1  \label{e1}
\end{equation}
The property space is therefor 2-dimensional. This, of course, imposes
limits on complexity of the simulations since the particles can never leave
their property domain; hence, the simulations are restricted to reproducing
only some features of complex evolutionary systems and are not aimed at full
emulation of any realistic complex evolution. The property values change due
to competitive mixing and random mutations. In order to be a winner,
particle $p$ must have smaller $y^{(i)}$ in majority of the properties $%
i=1,2,3$. The competition rules are strongly intransitive and intransitive
triplets 
\begin{equation}
\mathbf{y}_{p}\mathbf{\prec y}_{r}\mathbf{\prec y}_{q}\mathbf{\prec y}_{p}
\label{e2}
\end{equation}
can be found in the vicinity of any point in the property space. All
mutations are generated through the same function producing random values
uniformly distributed in the property domain. In the presented images, the
three particle properties are represented as a combination of primary
colours red, green and blue. The physical domain is mapped into a
rectangular box by the erf(...) functions. The details of the numerical
simulations can be found in Ref. \cite{K-PS2010}. Although simulations
display qualitatively different types of behaviour at different time
periods, all parameters of the simulation are kept constant during the whole
of the Long Run.\bigskip

The author thanks D.A. Klimenko and K. Slaughter
for contributing to the code and for running the simulations.

\pagebreak

\section{The property triangle}

\begin{figure}[h]
\label{fig1}
\par
\begin{center}
%
%
\par
{\includegraphics[width=12cm,height=12cm]{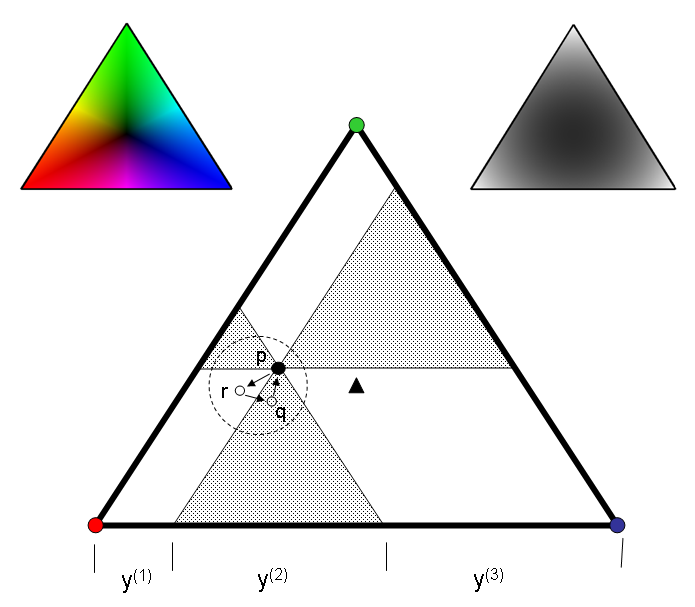} }
\end{center}
\caption{The property domain; left: representation of the properties by
colours, right: relative ranking with respect to the uniform distribution
(higher ranks are brighter), bottom: dominating (white) and dominated (grey)
regions for particle $p$. }
\end{figure}

The property domain represents a triangle with vertices corresponding to
pure colours: red $(1,0,0)$, green $(0,1,0)$ and blue $(0,0,1).$ The
intermediate colours are shown in the left figure. The competition rules are
selected so that each location $p$ dominates over grey areas and is
dominated by the white areas of the triangle. These rules are strongly
intransitive and intransitive triplets given by equation (\ref{e2}) can be
found in the vicinity of every location. One of these triplets is shown in
the figure with arrows directed from the losers to the winners.

The strongest points are the vertices of the triangle while the centre of
the triangle $(1/3,1/3,1/3),$ which is marked by the small black triangle,
is the weakest point (if measured with respect to the uniform distribution
-- the corresponding relative ranking is shown in the right figure).

\pagebreak

\section{Graphic representation}

\begin{figure}[h]
\label{fig2}
\par
\begin{center}
{\includegraphics[width=12cm,height=7cm]{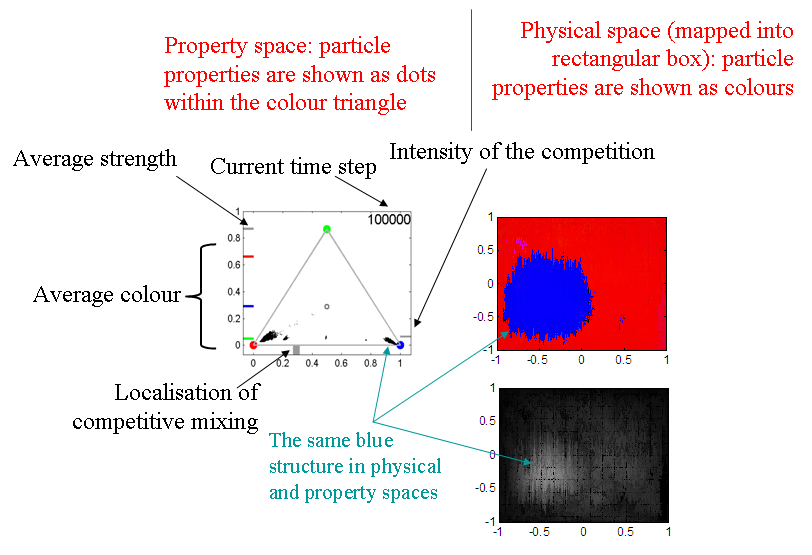} }
\end{center}
\caption{Representation of the property and physical domains.}
\end{figure}

In the physical domain particle properties shown at their physical locations
and interpreted as a combination of the primary colours: red, green and
blue. Brighter colours correspond to properties that, on average, tend to
perform better in competition. The weaker sets of properties are dark. For
an average human eye, it is difficult to distinguish finer grades of
brightness and colours at the same time. The black and white image is added
to show existence of structure within the spots. Some of the figures
presented in this supplement are darkened to visualise these structures: the
leading group of particles tend to be located near the centre of the spots.
In the property space, these structures resemble pyramids with the leading
particles located at the top of the pyramids and low ranked particles
occupying the bottoms. Note that gradations \ of brightness may become
indistinguishable in printed copies (depending on the printer used). The
darkened format, however, makes distinguishing colours more difficult and
most of the images are shown with bright colours, which hide existence of
structure within each spot. The vertical darker lines do not correspond to
any structure and are artifacts of the imaging algorithm.

In the property domain, the particle properties are shown irrespective of
their physical locations by small black dots within the property triangle.
These snapshots show additional average information: average values of $%
y^{(i)},$ $i=1,2,3,$ the intensity of competition, current time step, etc.
The intensity of competition is defined as the average readjustment of
particle properties due to competition 
\begin{equation}
\Xi =\left\langle \left| \mathbf{\acute{y}-y}\right| \right\rangle ,
\label{e3}
\end{equation}
where the average is calculated over all losing particles while $\mathbf{y}$
and $\mathbf{\acute{y}}$ represent the property values before and after a
time step.

\pagebreak

\section{Simulation snapshots}

\begin{figure}[h]
\label{fig3}
\par
\begin{center}
{\includegraphics[width=12cm,height=8cm]{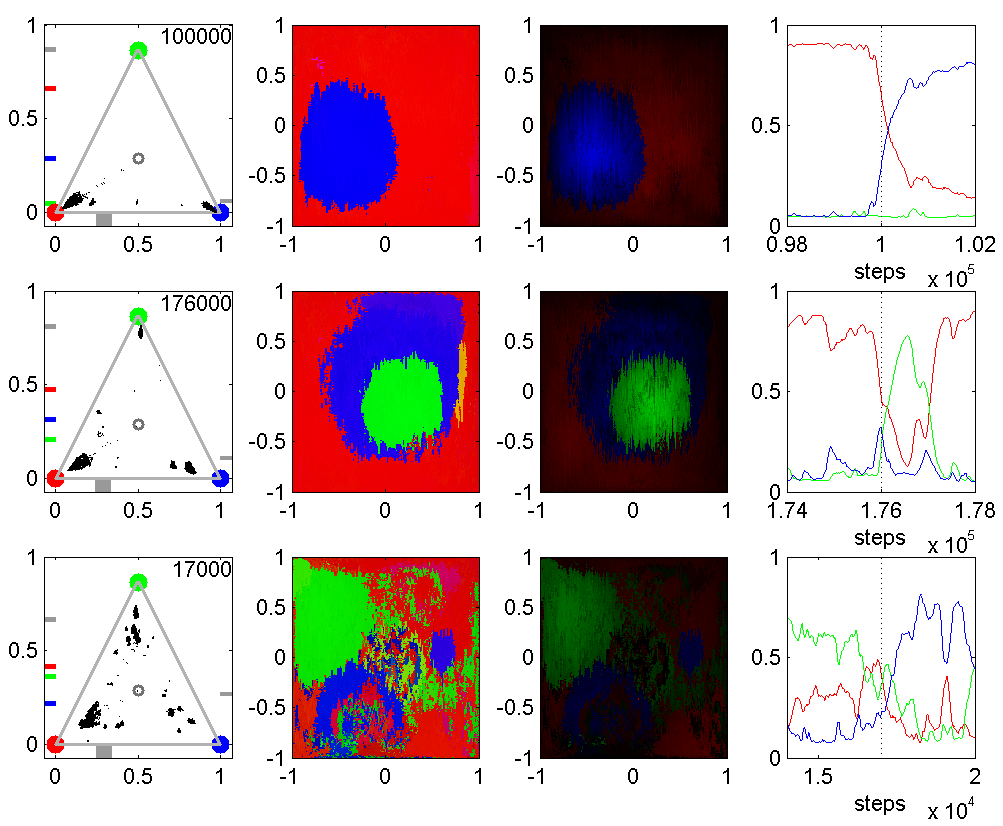} }
\end{center}
\caption{Property domain snapshot (left column), physical domain snapshot
(left middle column), physical domain snapshot with darkened colours (right
middle column) and average properties \TEXTsymbol{<}$y^{(i)}$\TEXTsymbol{>}, 
$i=1,2,3,$ (right column). The vertical dotted line in the right column
indicates the time step of the snapshot.}
\end{figure}

\begin{itemize}
\item  \textbf{Top row.}{\tiny \ }{\small A very strong blue spot challenges
the persistent dominance of the red colour and proceeds to take control over
the domain. The darker image shows the structure of the spots that takes a
pyramid-like shape in the property triangle. Formation of spots and
existence of hierarchy within the spots reduce the competition intensity }$%
\Xi ,${\small \ since the particles compete with other particles that have
the same type of the colour (within the same spot) and similar strength
(having similar places in the hierarchy) while essential struggle is limited
to the frontlines separating different spots. This, combined with the common
fate of particles within the same spot (i.e. a high likelihood of a joint
success or a joint failure), represents \textit{competitive cooperation}.}

\item  \textbf{Middle row.}{\tiny \ }{\small Although all three primary
colours (red, green and blue) generally have the same strength, the specific
properties of the leading particles at this time step make blue strong
against red and green strong against blue (the reader is referred to the
regions of domination shown in Figure 1 for an arbitrary particle }$p$%
{\small ). Thus, green is using blue to capture the central regions
previously controlled by red. When green easily eliminates blue, it is
well-positioned to face a more difficult opponent (the red) and, after a
struggle, almost manages to defeat this opponent. At the brink of its
complete elimination, the red responds by a successful mutation and recovers.%
}

\item  \textbf{Bottom row.}{\tiny \ }{\small Green global dominance is
subject to \textit{competitive degradation} and declines. The weakening
green has to face an increasing number of challengers. The struggle becomes
chaotic, distributed over interior areas and not limited to the frontlines
(one may call this ``total competition''). This dramatically increases the
intensity of competition and significantly reduces the level of competitive
cooperation (but even ``total competition'' does not eliminate cooperation
completely). Green responds by a successful mutation bringing new
leadership, which achieves a noticeable success in restoring green rule over
some regions in the upper left corner of the physical domain but the
insufficient strength of the green structure and its poor strategic
positioning (note that in this case green is stronger against red than
against blue) makes the successful completion of this task unlikely.}
\end{itemize}

\pagebreak

\section{The Leaping Cycle}

\begin{figure}[h]
\label{fig4}
\par
\begin{center}
{\includegraphics[width=12cm,height=10cm]{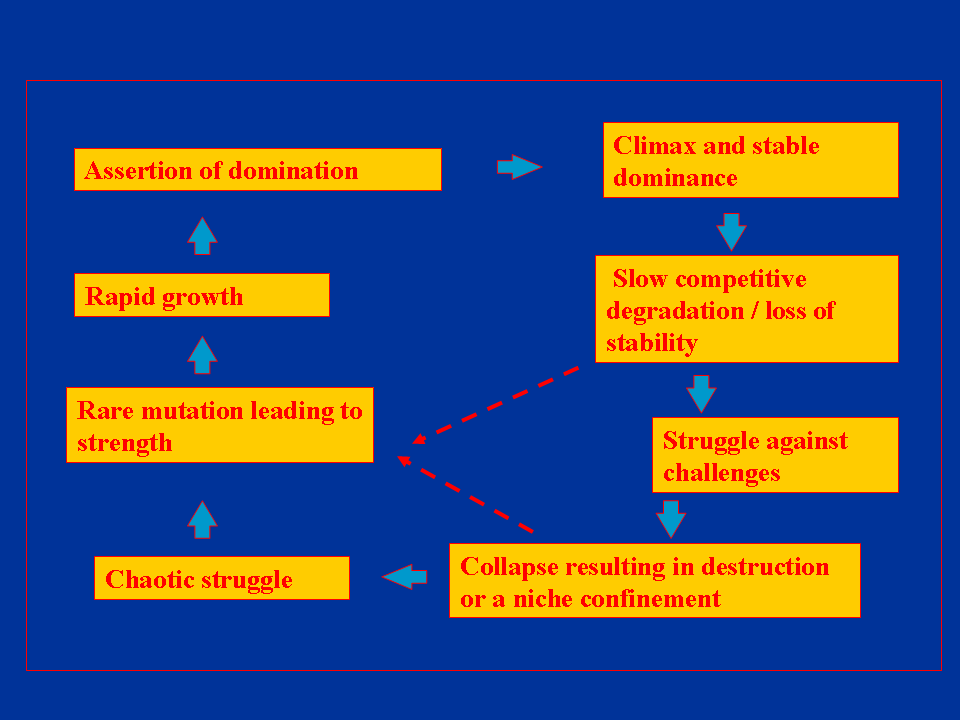} }
\end{center}
\caption{Schematic of the leaping cycle. Dashed arrows idicate possible
alternatives.}
\end{figure}

The \textit{Leaping cycle,} which is the generic form of the cycle of
emergence, rapid growth, stable dominance, decline and, possibly, collapse,
is caused by intransitivity of the competition. In intransitive
competitions, dominant structures tend to adapt to existing competitors but
may become unstable with respect to potential challenges not currently
present in the system. Competitive strength can be eroded by \textit{%
competitive degradation} resulting in decline followed by collapse or
marginalised existence. As a new strong structure emerges and leaps into a
dominant position, the leaping cycle repeats itself. If there is no strong
competitor present, the old dominant structure may recover or may collapse.
In the latter case the competition may become chaotic and the chaotic
struggle continues until a new structure of sufficient strength emerges. Red
arrows in the figure show alternative possibilities involving bypassing some
of the stages. 

The Leaping cycle represents a generic form of the cycle (i.e. examined in
the context of abstract competition). As discussed in the main text,
analogues of this cycle are known in different disciplines under different
names. Note that the leaping cycle differs from ordinary cycles, which
result in recurrent repetition of the same or similar states.

\pagebreak

\section{The Long Run}

\begin{figure}[h]
\label{fig5}
\par
\begin{center}
{\includegraphics[width=12cm,height=9cm]{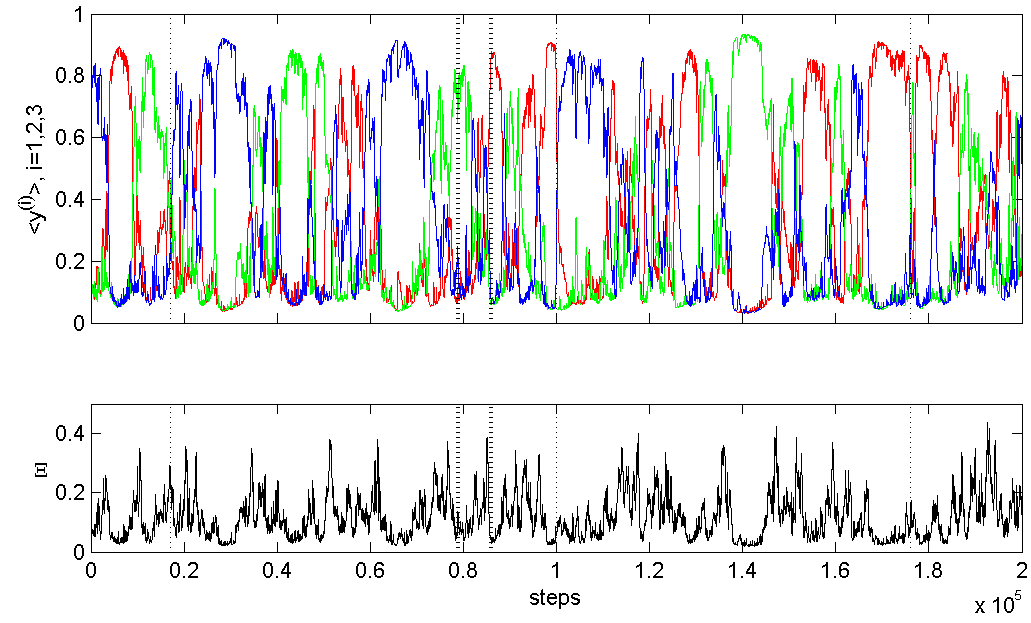} }
\end{center}
\caption{Average values of the properties $\left\langle y^{(i)}\right\rangle 
$, $i=1,2,3$ (top figure) and the competition intensity $\Xi$ (bottom
figure) for the Long Run of 200 000 time steps. The vertical dotted lines
indicate locations of the snapshots shown in Figure \ref{fig3}, while thick
dotted lines correspond to the time span of the video fragment.}
\end{figure}

The Long Run history clearly shows persistence of the \textit{leaping cycle}
in the simulations. Dominance of one of the colours tend to be gradually
reduced until a new dominant colour emerges. The periods with a dominating
colour (``empires'') are interchanged with periods of chaotic struggle where
emerging structures are insufficiently strong to survive for a long time.
The chaotic periods are full of various events and are much more interesting
to watch than boring imperial periods, where very few changes take place.
The chaotic struggle is, of course, not fully random and is controlled by a
combination of a more adverse competition between the structures and \textit{%
competitive cooperation }within the structures: interdependencies between
particles and structures still exist but have relatively short survival
times.

The ``empires'' are subject to \textit{competitive degradation} and do not
last forever. Competition of structures of different colours and similar
strengths tends to disintegrate into chaotic struggles distributed over
areas (i.e. ``total competition'') when no clear winner can emerge from the
competition. A stronger structure tends to preserve its spot-like shape
while taking over the areas previously controlled by noticeably weaker
structures.

\pagebreak

\section{Video Fragment for Evolution of a Complex Competitive System}

\begin{figure}[h]
\label{fig6}
\par
\begin{center}
{\includegraphics[width=11cm,height=7cm]{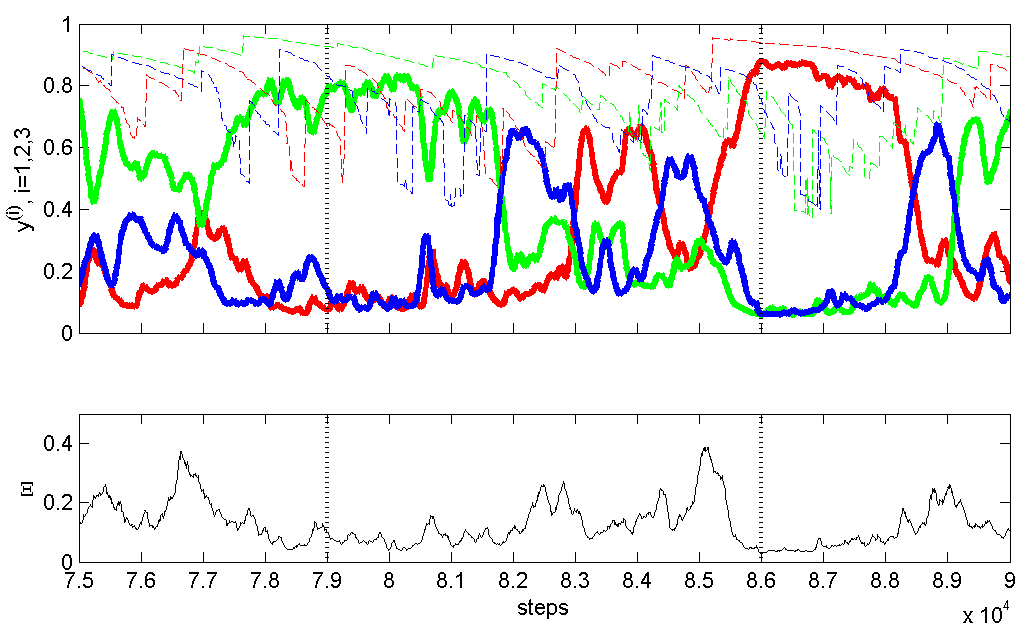} }
\end{center}
\caption{ Average $\left\langle y^{(i)}\right\rangle $ (top figure, thick
solid lines) and maximal $\func{max}(y^{(i)})$ (top figure, dashed lines)
values $i=1,2,3$ for the time span of the video (which is shown by the
vertical dotted lines). The bottom figure shows the competition intensity $%
\Xi $ specified by (\ref{e3}).}
\end{figure}

\textbf{Degradation.} {\small The video starts from the state of prominent
dominance of the green colour (}$i=2${\small ). \ The green dominance
gradually weakens due to \textit{competitive degradation}, which is
indicated by reduction of }$\max (y^{(2)})${\small \ further away from its
strongest possible value of }$y^{(2)}=1${\small . Between 80 and 81 thousand
steps, green dominance faces an essential crisis but, aided by a successful
reformatory mutation, which increases }$\max (y^{(2)})${\small , the green
manages to restore its control over the domain. Reformatory mutations may
temporarily reverse some effects of competitive degradation but the
degradation always resumes. The further reduction of }$\max (y^{(2)})$%
{\small \ is inevitably followed by the loss of the territory and decrease
of }$\left\langle y^{(2)}\right\rangle ${\small . \ While the green
initially manages to defend its dominance against competitors, this task
becomes increasingly difficult as the green power declines. The competition
intensity }$\Xi ${\small , which was low under green dominance, increases.}

\textbf{Chaotic struggle.} {\small Finally, the green collapses at 82
thousand steps but none of the winners is strong enough to establish
themselves as a new dominant power. The struggle becomes chaotic and, at
times, slides into dispersed ``total competition'' --- at these points the
competition intensity becomes extremely high. Structures emerge for a short
time, when local conditions are favourable, only to disappear without a
trace.}

\textbf{Emergence.} {\small A series of successful mutations leads to a new
red strength that easily takes over chaotic regions and then overcomes the
remaining blue resistance. The complete dominance of red colour (}$i=1$%
{\small ) is then promptly established, just before 86 thousand steps. This
completes the \textit{leaping cycle}. The competition intensity drops to its
minimum. It is interesting that as victorious red still expands its
influence over the whole domain, the sneaky competitive degradation is
already present and reduces }$\max (y^{(1)}).${\small \ While the red
dominance at the end of the video seems unshakable, its stealthy decline has
already began and soon will become more and more visible. The red dominance
will not last forever and the \textit{leaping cycle} will soon resume its
pace.}\bigskip

The video is sped up and reduced in size by displaying only each fifth frame
of the simulations. The event-rich middle section of the video needs to be
watched at a reduced frame rate to see the details of the evolution.

\pagebreak


%% file: ins1sup.tex
{\Large \it Appendix}
\begin{center}
{{\LARGE \bf Example of complex behaviour \\ in competitive systems } \\ 
\bigskip
\bigskip
{\small \bf This appendix is a part of the electronic supplement for the article \\ Phil. Trans. R. Soc. A 2013, vol.371, No 1982, 20120244 \\  }
}

\end{center}

\bigskip
\bigskip

This appendix presents the results of numerical simulations of abstract competition based on a fixed set 
of elementary competition rules. The system under consideration is one of the simplest possible systems
with strongly intransitive competition rules and competitive mixing localised in physical space 
(i.e. this is one of the simplest possible representatives of complex competitive systems). 
The simulations display a number of features associated with complex behaviour including competitive cooperation and leaping cycle. 
Please note that these simulations are generic 
and are not intended to model any specific process or event. 
The supplement includes seven figures, brief explanations and one video file (7.5MB, mp4, H.264 encoded).

\begin{figure}[h]
\label{fig0}
\par
\begin{center}
%
%
\par
{\includegraphics[width=12cm,height=6cm]{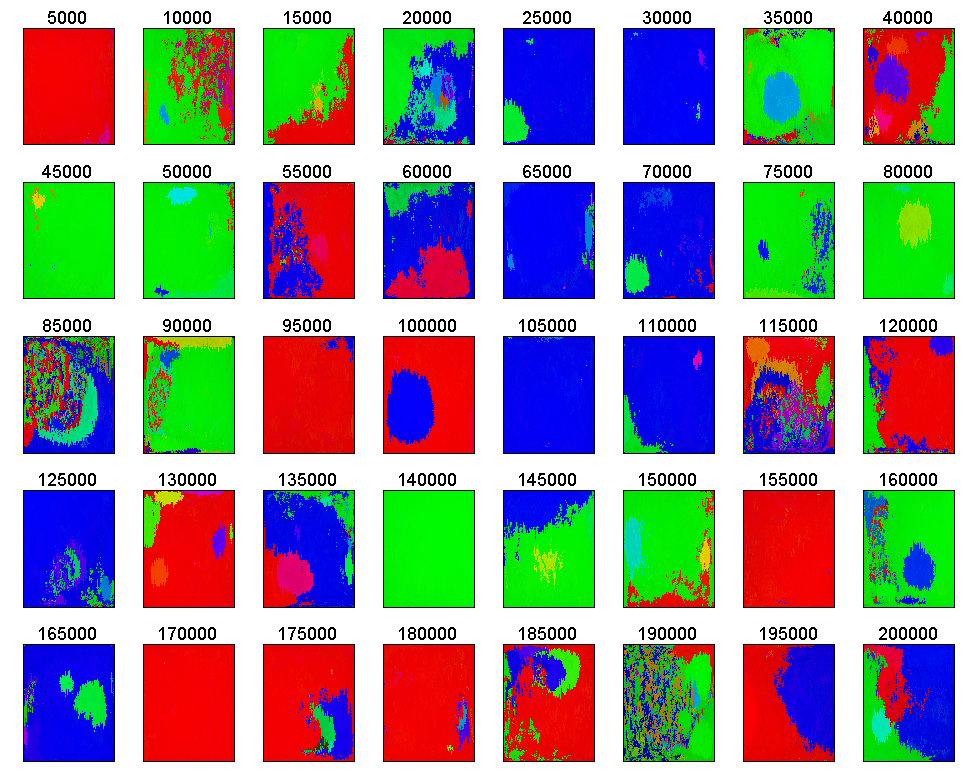} }
\end{center}
\end{figure}



%% file: paperFT3.bbl
\begin{thebibliography}{60}
\providecommand{\natexlab}[1]{#1}
\providecommand{\url}[1]{\texttt{#1}}
\expandafter\ifx\csname urlstyle\endcsname\relax
  \providecommand{\doi}[1]{doi: #1}\else
  \providecommand{\doi}{doi: \begingroup \urlstyle{rm}\Url}\fi

\bibitem[Schr\"{o}dinger(1944)]{Life1944}
E.~Schr\"{o}dinger.
\newblock \emph{What is Life - the Physical Aspect of the Living Cell}.
\newblock Cambridge University Press, Cambridge, UK, 1944.

\bibitem[Heylighen et~al.(1999)Heylighen, Bollen, and Riegler]{CompEvol1999}
F.~Heylighen, J.~Bollen, and A.~Riegler.
\newblock \emph{The evolution of complexity}, volume~8.
\newblock Kluwer, Dordrecht, 1999.

\bibitem[Nicolis and Prigogine(1977)]{Prig1977}
G.~Nicolis and I.~Prigogine.
\newblock \emph{Self-organization in nonequilibrium systems: from dissipative
  structures to order through fluctuations}.
\newblock Wiley, New York, 1977.

\bibitem[Sherrington(2010)]{ComPT2010}
D.~Sherrington.
\newblock Physics and complexity.
\newblock \emph{Philosophical Transactions of the Royal Society A:
  Mathematical, Physical and Engineering Sciences}, 368\penalty0
  (1914):\penalty0 1175--1189, 2010.

\bibitem[Devine(2009)]{AlgEntrop2009}
S.~Devine.
\newblock The insights of algorithmic entropy.
\newblock \emph{Entropy}, 11\penalty0 (1):\penalty0 85--110, 2009.

\bibitem[Gell-Mann(1994)]{G-M1994}
M.~Gell-Mann.
\newblock \emph{The quark and the jaguar}.
\newblock W.H.Freemen and Co., N.-Y., 1994.

\bibitem[Holland(2006)]{CAS2006}
J.H. Holland.
\newblock Studying complex adaptive systems.
\newblock \emph{J. of Syst. Sci. and Complexity}, 19\penalty0 (1):\penalty0
  1--8, 2006.

\bibitem[Klimenko(2010{\natexlab{a}})]{K-PS2010}
A.~Y. Klimenko.
\newblock Conservative and competitive mixing and their applications.
\newblock \emph{Physica Scripta}, T142:\penalty0 014054, 2010{\natexlab{a}}.

\bibitem[Klimenko and Pope.(2012)]{KP2012}
A.~Y. Klimenko and S.~B. Pope.
\newblock Propagation speed of combustion and invasion waves in stochastic
  simulations with competitive mixing.
\newblock \emph{Combustion Theory and Modelling}, 16\penalty0 (4):\penalty0
  679--714, 2012.
\newblock \doi{10.1080/13647830.2011.647091}.

\bibitem[Klimenko(2012)]{K-PS2012}
A.~Y. Klimenko.
\newblock Mixing, entropy and competition.
\newblock \emph{Physica Scripta}, 85:\penalty0 068201, 2012.

\bibitem[Pope(1985)]{Pope85}
S.~B. Pope.
\newblock Pdf methods for turbulent reactive flows.
\newblock \emph{Prog. Energy Combust. Sci.}, 11:\penalty0 119--192, 1985.

\bibitem[Williams(1985)]{Williams85}
F.~A. Williams.
\newblock \emph{Combustion Theory}.
\newblock Addison-Wesley, Reading, MA, 2nd edition, 1985.

\bibitem[Kuznetsov and Sabelnikov(1990)]{Kuznetsov-S-90}
V.~R. Kuznetsov and V.~A. Sabelnikov.
\newblock \emph{Turbulence and Combustion}.
\newblock Hemisphere, New York, 1990.

\bibitem[Warnatz et~al.(1996)Warnatz, Maas, and Dibble]{Maas1996}
J.~Warnatz, U.~Maas, and R.~W. Dibble.
\newblock \emph{Combustion: physical and chemical fundamentals, modelling and
  simulation, experiments, pollutant formation}.
\newblock Springer-Verlag, Berlin, 1996.

\bibitem[Klimenko and Bilger(1999)]{KB99}
A.~Y. Klimenko and R.~W. Bilger.
\newblock Conditional moment closure for turbulent combustion.
\newblock \emph{Prog. Energy Combust. Sci.}, 25:\penalty0 595--687, 1999.

\bibitem[Pope(2000)]{Pope-00}
S.~B. Pope.
\newblock \emph{Turbulent Flows}.
\newblock Cambridge University Press, Cambridge, 2000.

\bibitem[Peters(2000)]{Peters-00}
N.~Peters.
\newblock \emph{Turbulent Combustion}.
\newblock Cambridge University Press, 2000.

\bibitem[Fox(2003)]{Fox-2003}
R.~Fox.
\newblock \emph{Computational Models for Turbulent Reacting Flows}.
\newblock Cambridge University Press, Cambridge, 2003.

\bibitem[Heinz(2003)]{Heinz-2003}
S.~Heinz.
\newblock \emph{Statistical Mechanics of Turbulent Flows}.
\newblock Springer, Berlin, 2003.

\bibitem[Pitsch(2006)]{Pitsch2006}
H.~Pitsch.
\newblock {Large-Eddy} simulations of turbulent combustion.
\newblock \emph{Annu. Rev. Fluid Mech.}, 38:\penalty0 453--482, 2006.

\bibitem[Haworth(2010)]{Haworth2009}
D.~Haworth.
\newblock Progress in probability density function methods for turbulent
  reacting flows.
\newblock \emph{Prog. Energy Combust. Sci.}, 36:\penalty0 168--259, 2010.

\bibitem[Echekki and Mastorakos(2011)]{book2011}
T.~Echekki and E.~Mastorakos.
\newblock \emph{Turbulent Combustion Modelling. Advances, New Trends and
  Perspectives.}
\newblock Springer, 2011.

\bibitem[Klimenko and Cleary(2010)]{KC-PopeFTC}
A.~Y. Klimenko and M.~J. Cleary.
\newblock Convergence to a model in sparse-lagrangian fdf simulations.
\newblock \emph{Flow, Turbulence and Combustion (Special issue dedicated to
  S.B.Pope)}, 85:\penalty0 567–591, 2010.

\bibitem[Caratheodory(1909)]{Cara1909}
C.~Caratheodory.
\newblock Studies in the foundation of thermodynamics.
\newblock \emph{Math. Ann.}, 67:\penalty0 355–386, 1909.

\bibitem[Lieb and Yngvason(2003)]{EntOrd2003}
E.~H. Lieb and J.~Yngvason.
\newblock The entropy of classical thermodynamics.
\newblock In A.~Greven, G.~Keller, and G.~Warnecke, editors, \emph{Entropy},
  chapter~8, pages 147--195. Princeton University Press, 2003.

\bibitem[Kindermann and Snell(1980)]{GibbsMeasure80}
R.~Kindermann and J.~L. Snell.
\newblock \emph{Markov random fields and their applications}.
\newblock AMS, 1980.

\bibitem[Evans and Searles(2002)]{Ftheor2002}
D.J. Evans and D.J. Searles.
\newblock The fluctuation theorem.
\newblock \emph{Advances in Physics}, 51\penalty0 (7):\penalty0 1529--1585,
  2002.

\bibitem[Glansdorff and Prigogine(1971)]{PrigGlan}
P.~Glansdorff and I.~Prigogine.
\newblock \emph{Thermodynamic theory of structure, stability and fluctuations}.
\newblock Wiley-Interscience, London ; New York, 1971.

\bibitem[Paltridge(1979)]{MAP1979}
G.~W. Paltridge.
\newblock Climate and thermodynamic systems of maximum dissipation.
\newblock \emph{Nature}, 279\penalty0 (5714):\penalty0 630--631, 1979.

\bibitem[Ziegler(1983)]{Ziegler1983}
H.~Ziegler.
\newblock \emph{An introduction to thermomechanics}, volume 21.
\newblock North-Holland, Amsterdam ; New York, 1983.

\bibitem[Martyushev and Seleznev(2006)]{MEP2006}
L.~M. Martyushev and V.~D. Seleznev.
\newblock Maximum entropy production principle in physics, chemistry and
  biology.
\newblock \emph{Physics Reports}, 426\penalty0 (1):\penalty0 1--45, 2006.

\bibitem[Mehta(1999)]{Mehta1999}
G.~B. Mehta.
\newblock Preference and utility.
\newblock In S.~Barbera, P.~Hammond, and C.~Seidl, editors, \emph{Handbook of
  utility theory}, pages 1–--47. Kluwer, Boston, 1999.

\bibitem[Perez(2010)]{Perez2010}
C.~Perez.
\newblock Technological revolutions and techno-economic paradigms.
\newblock \emph{Camb. J. Econ.}, 34\penalty0 (1):\penalty0 185--202, 2010.

\bibitem[Eigen(1971)]{Eigen1971}
M.~Eigen.
\newblock Selforganization of matter and the evolution of biological
  macromolecules.
\newblock \emph{Naturwissenschaften}, 58:\penalty0 465--523, 1971.

\bibitem[{de Condorcet}(1785)]{Cond1785}
N.~{de Condorcet}.
\newblock \emph{Essay on the Application of Analysis to the Probability of
  Majority Decisions}.
\newblock De L'imprimerie Royale., Paris, 1785.

\bibitem[Monin and Yaglom(1975)]{MoninY1975}
A.~S. Monin and A.~M. Yaglom.
\newblock \emph{Statistical Fluid Mechanics}.
\newblock M.I.T. Press, Cambridge, MA, 1975.

\bibitem[Klimenko(2010{\natexlab{b}})]{KlimIMCIC2010}
A.~Y. Klimenko.
\newblock Computer simulations of abstract competition.
\newblock In \emph{Proceedings of International Conference on Complexity,
  Informatics and Cybernetics (IMCIC)}, volume~1, pages 97--102,
  2010{\natexlab{b}}.

\bibitem[Debreu(1954)]{Debreu1954}
G.~Debreu.
\newblock Presentation of a preference odering by a numerical function.
\newblock In R.~M. Thrall, C.~H. Coombs, and R.~L. Davis, editors,
  \emph{Decision process}, pages 159--165. J.Wiley and Sons, 1954.

\bibitem[Pope and Anand(1985)]{Pope-A-85}
S.~B. Pope and M.~S. Anand.
\newblock Flamelet and distributed combustion in premixed turbulent flames.
\newblock In \emph{Proc. Combust. Inst}, volume~20, pages 403--410, Pittsburgh,
  1985. Combustion Institute.

\bibitem[Fisher(1937)]{Fisher1937}
R.~A. Fisher.
\newblock The wave of advance of advantageous genes.
\newblock \emph{Ann. Eugen.}, 7:\penalty0 355, 1937.

\bibitem[Kolmogorov et~al.(1937)Kolmogorov, Petrovsky, and Piskunov]{KPP1937}
A.~N. Kolmogorov, I.~G. Petrovsky, and N.~S. Piskunov.
\newblock A study of the equation of diffusion with increase in the quantity of
  matter, and its application to a biological problem.
\newblock \emph{Mosc. Univ. Bull. (A: Math. and Mech.)}, 1:\penalty0 1, 1937.

\bibitem[Bray et~al.(1984)Bray, Libby, and Moss]{Bray-L-M-84a}
K.~N.~C. Bray, P.~A. Libby, and J.~B. Moss.
\newblock Flamelet crossing frequencies and mean reaction rates in premixed
  turbulent combustion.
\newblock \emph{Combust. Sci. Technol.}, 41:\penalty0 143--172, 1984.

\bibitem[Mollison(1977)]{Mollison1977}
D.~Mollison.
\newblock Spatial contact models for ecological and epidemic spread.
\newblock \emph{J. R. Statist. Soc.}, B 39\penalty0 (3):\penalty0 283, 1977.

\bibitem[Dupree(1992)]{Dupree1992}
T.~H. Dupree.
\newblock Coarse-grain entropy in two-dimensional turbulence.
\newblock \emph{Physics of Fluids B: Plasma Physics}, 4\penalty0 (10):\penalty0
  3101, 1992.

\bibitem[Falkovich and Fouxon(2004)]{Falk2004}
G.~Falkovich and A.~Fouxon.
\newblock Entropy production and extraction in dynamical systems and
  turbulence.
\newblock \emph{New Journal of Physics}, 6:\penalty0 50--50, 2004.

\bibitem[Candeal et~al.(2001)Candeal, {De Miguel}, Indurain, and
  Mehta]{Mehta2001}
J.~C. Candeal, J.~R. {De Miguel}, E.~Indurain, and G.B. Mehta.
\newblock Utility and entropy.
\newblock \emph{Economic Theory}, 17\penalty0 (1):\penalty0 233--238, 2001.

\bibitem[Marshall et~al.(1982)Marshall, Webb, {Sepkoski,Jr}, and Raup]{GAI1982}
L.G. Marshall, S.~D. Webb, J.~J. {Sepkoski,Jr}, and D.~M. Raup.
\newblock Mammalian evolution and the great american interchange.
\newblock \emph{Science}, 215\penalty0 (4538):\penalty0 1351--1357, 1982.

\bibitem[Whitfield(2005)]{MAP2005}
J.~Whitfield.
\newblock Order out of chaos.
\newblock \emph{Nature}, 436\penalty0 (7053):\penalty0 905--907, 2005.

\bibitem[Tullock(1964)]{Intrans1964}
G.~Tullock.
\newblock The irrationality of intransitivity.
\newblock \emph{Oxford Economic Papers}, 16\penalty0 (3):\penalty0 401--406,
  1964.

\bibitem[Arrow(1951)]{Arrow}
K.~J. Arrow.
\newblock \emph{Social Choice and Individual Values}.
\newblock Yale University Press, USA, 1951.

\bibitem[Anand(1993)]{Intrans1993}
P.~Anand.
\newblock The philosophy of intransitive preference.
\newblock \emph{Economic Journal}, 103\penalty0 (417):\penalty0 337--346, 1993.

\bibitem[Makowski and Piotrowski(2011)]{Intrans2011}
M.~Makowski and E.~W. Piotrowski.
\newblock Decisions in elections -- transitive or intransitive quantum
  preferences.
\newblock \emph{Journal of Physics A: Mathematical and Theoretical},
  44\penalty0 (21):\penalty0 215303, 2011.

\bibitem[Boddy(2000)]{BioInt2000}
Lynne Boddy.
\newblock Interspecific combative interactions between wood-decaying
  basidiomycetes.
\newblock \emph{FEMS Microbiology Ecology}, 31\penalty0 (3):\penalty0 185--194,
  2000.

\bibitem[Bouyssou(2004)]{Rank2004}
D.~Bouyssou.
\newblock Monotonicity of 'ranking by choosing': A progress report.
\newblock \emph{Social Choice and Welfare}, 23\penalty0 (2):\penalty0 249--273,
  2004.

\bibitem[Kondratiev(1984)]{Kond1925}
N.~D. Kondratiev.
\newblock \emph{The Long Wave Cycle (translation from Russian "The Major
  Economic Cycles", 1925)}.
\newblock Richardson and Snyder, NY, 1984.

\bibitem[Batchelor(1953)]{Batchelor-53}
G.~K. Batchelor.
\newblock \emph{The theory of homogeneous turbulence}.
\newblock Cambridge University Press, Cambridge, 1953.

\bibitem[Tsinober(2009)]{Tsinober2009}
A.~Tsinober.
\newblock \emph{An Informal Conceptual Introduction to Turbulence}.
\newblock Springer, London, New York, 2009.

\bibitem[Sommeria(2005)]{Som2005}
J.~Sommeria.
\newblock Entropy production in turbulent mixing.
\newblock In M~Lesieur, A.~Yaglom, and F.~David, editors, \emph{New trends in
  turbulence}, pages 79--91. Springer Berlin Heidelberg, Berlin, Heidelberg,
  2005.

\bibitem[Borisov et~al.(1989)Borisov, Kuznetsov, and Shedogubov]{Kuz1989}
A.~G. Borisov, V.~R. Kuznetsov, and Y.~M. Shedogubov.
\newblock Influence of linear and nonlinear effects on the large-scale
  structure of turbulent jet-type shear flows.
\newblock \emph{Fluid Dynamics}, 24\penalty0 (4):\penalty0 525--533, 1989.

\bibitem[Gyftopoulos and Beretta(1991)]{Beretta1991}
E.P. Gyftopoulos and G.P. Beretta.
\newblock \emph{Thermodynamics. Foundations and Applications}.
\newblock Macmillan, N.Y., USA, 1991.

\end{thebibliography}
